\documentclass[conference]{IEEEtran}

\usepackage[latin1]{inputenc}
\usepackage{amsmath}
\usepackage{amsfonts}
\usepackage{amssymb}
\usepackage{graphicx}
\usepackage{mathcomSTEv4}
\usepackage{tikz,multirow}
\usepackage{pgfplots}
\usepackage{algorithm}
\usepackage[noend]{algpseudocode}

\def\BState{\State\hskip-\ALG@thistlm}

\newcommand{\Lmf}{\mathfrak{L}}

\tikzstyle{joint} = [draw, circle, minimum size=.1em]
\usetikzlibrary{positioning}
\usetikzlibrary{arrows,fit}
\usetikzlibrary{shapes}
\usetikzlibrary{arrows.meta}
\usetikzlibrary{decorations.pathreplacing}
\tikzstyle{int}=[draw, fill=blue!10,thick,minimum height = .75 cm, minimum width=1cm]
\tikzstyle{rbox}=[draw, rounded corners=5 pt,thick ]
\tikzstyle{sum}=[circle, fill=blue!10, draw=black,line width=.5 pt,minimum size = 0.05 cm, thin ]
\tikzstyle{joint} = [draw, circle, minimum size=1em]
\usetikzlibrary{shapes,arrows,backgrounds,positioning}
\usetikzlibrary{positioning,intersections,decorations.pathreplacing}
\tikzstyle{cw}= [fill=blue!10, draw , rounded corners = 1 ex,minimum height = 1cm]
\tikzstyle{joint} = [draw, circle, minimum size=1em]
\tikzstyle{check} = [draw,  minimum size=1em]

\allowdisplaybreaks

\title{
	LDPC Coded Multiuser Shaping for the Gaussian Multiple Access Channel
}

\author{
	\IEEEauthorblockN{Alexios Balatsoukas-Stimming}
	\IEEEauthorblockA{
		EPFL, Switzerland\\
		{alexios.balatsoukas@epfl.ch}
	}
	\and
	\IEEEauthorblockN{Stefano Rini}
	\IEEEauthorblockA{
		NCTU, Taiwan\\
		{stefano@nctu.edu.tw}
	}
	\and
	\IEEEauthorblockN{J\"org Kliewer}
	\IEEEauthorblockA{
		NJIT, USA\\
		{jkliewer@njit.edu}
	}
}

\begin{document}
	\maketitle
	\begin{abstract}
	The joint design of input constellation and low-density parity-check (LDPC) codes to approach the symmetric capacity of the two-user Gaussian multiple access channel is studied. More specifically, multilevel coding is employed at each user to construct a high-order input constellation and the constellations of the users are jointly designed so as to maximize the multiuser shaping gain. At the receiver, each layer of the multilevel coding is jointly decoded among users, while successive cancellation is employed across layers. The LDPC code employed by each user in each layer is designed using EXIT charts to support joint decoding among users for the prescribed per-layer rate and SNR. Numerical simulations are provided to validate the proposed constellation and LDPC code designs.
	\end{abstract}

	\begin{IEEEkeywords}
		Gaussian Multiple Access Channel; LDPC codes; Joint decoding; Multilevel coding; Constellation shaping.
	\end{IEEEkeywords}

\section{Introduction}
With the advent of machine-type communication, the problem of uncoordinated channel access by multiple users has once again attracted the attention of both researchers and practioners.
As the number of users in wireless networks and the transmission rates are ever-increasing, multiuser medium access invariably emerges as the bottleneck to the network performance.
A natural model to investigate multiuser medium access is the Gaussian multiple access channel (GMAC): in this channel, the receiver output is obtained as the sum of the input at the transmitters plus additive Gaussian noise.

The design of low-density parity-check (LDPC) codes for the GMAC was first considered in \cite{amraoui2002achieving}.
The authors of \cite{roumy2007characterization} introduce the concept of a ``MAC node'' for the factor graph when describing the decoding of LDPC codes for the GMAC.
This node is a third type of node, together with variable and check nodes, which receives the channel output and the bit-reliability for a symbol of one transmitter and produces the bit-reliability of a symbol of the other transmitter.
In \cite{wu2004best}, the authors propose a soft demapping method for multilevel modulation on the GMAC based on LDPC codes and investigate the role of symbol mapping in this setting.
Spatially coupled codes for the binary adder channel with erasures are studied in \cite{kudekar2011spatially} where it is shown that threshold saturation as in the point-to-point erasure channel also occurs in this model.
In \cite{yedla2011universal} spatially-coupled codes for the GMAC are studied, and is shown that threshold saturation occurs for the  joint decoding of two codewords: this result naturally leads to the design of codes which are universal with respect to the channel parameters.
In \cite{liang2017joint}, the authors study the construction of codes for the binary input MAC channel in which the variable node distribution is jointly designed among the two users, thus specifying the portion of nodes of a certain degree from one user that collide with a portion of nodes of a certain degree from the other user.
More recently, spurred by the demand of multiuser superposition transmission (MUST) non-orthogonal multiple access (NOMA) transmission schemes for machine-type communication in LTE and 5G, the unsourced MAC channel has emerged as a model of considerable interest.
This model corresponds to a MAC channel with a large number of total users but where only a small fraction is active.
For this model, a finite blocklength analysis is presented in \cite{ordentlich2017low}.
In \cite{vem2017user} a practical code design using LDPC codes is presented: in this design users employ IRA codes and different users are identified by the permutation employed in the code. A concatenated LDPC-repetition scheme that is effective at very low per-user rates is described in~\cite{Wang2018}.
We note that existing schemes focus almost exclusively on binary inputs.

\subsubsection*{Contribution}
In this work, we propose a novel joint transmit constellation and error-correcting code design for the two-user symbol-synchronous GMAC with $L$-ary inputs.
In particular, we introduce a novel design principle which we denote by the term \emph{multiuser constellation shaping}: this corresponds to the shaping of the sum constellation observed at the receiver, obtained through the over-the-air sum of the constellations of the single users.
We show how multiuser constellation shaping allows one to turn the geometric shaping of the constellation of the single users results into the probabilistic and geometric shaping of the sum constellation at the receiver.
More specifically, our code design is based on irregular LDPC codes while the constellation design relies on three components: (i) multilevel coding to attain high transmission rates (ii) multiuser constellation shaping to harness shaping gain in the receiver sum-constellation, and (iii) joint per-layer decoding across users to improve scalability.
We employ EXIT charts with a Gaussian approximation (GA) of the message distributions to investigate the convergence of the decoding process and derive a linear programming (LP) technique for joint per-layer, across-users code design.
%`
We design of the user constellations so that the resulting sum constellation yields the largest multiuser shaping gain for a given signal-to-noise ratio (SNR) under the condition of successful decoding.
Simulation results demonstrate the sum-rate improvements that can be obtained through the proposed joint design.

\section{Channel Model and Coding Scheme Description}\label{sec:Channel model and Code description}
\begin{figure}
	\centering
	\begin{tikzpicture}[node distance=2.25cm,auto,>=latex]
	\node at (0,0) (source1) {$W_1$};
	\node [int] (enc1) [right of = source1, node distance = 1.5 cm]{Enc. 1};
	\node at (0,-2) (source2) {$W_2$};
	\node [int] (enc2) [right of = source2, node distance = 1.5 cm]{Enc. 2};
	\node at (6,-1) [int] (dec) [node distance =2.5  cm]{Dec.};
	\node (sink) [right of = dec, node distance = 2 cm] {$\Wh_1,\Wh_2$};
	%
	%\node (Pyx1x2v1) [joint,left of = dec, node distance = 2.5 cm]{};
	\node (Pyx1x2) [rbox,left of = dec, node distance = 3 cm]{+};
	%
	%\node (n) [joint,left of = Pyx1x2, node distance = 2.5 cm]{};
	\node (n1) [rbox,right of = Pyx1x2, node distance = 1 cm]{+};

	\node (noise) [above of = n1, node distance = 1 cm]{$Z^n$};

	\draw[->,line width=1.5 pt] (enc1) -| node[above]{$X_1^n$} (Pyx1x2);
	\draw[->,line width=1.5 pt] (enc2) -| node[below]{$X_2^n$} (Pyx1x2);
	\draw[->,line width=1.5 pt] (Pyx1x2) -- (n1);
	\draw[->,line width=1.5 pt] (n1) -- node[below]{$Y^n$} (dec);
	\draw[->,line width=1 pt] (source1) -- (enc1);
	\draw[->,line width=1 pt] (source2) -- (enc2);
	\draw[->,line width=1 pt] (dec) -- (sink);
	\draw[->,line width=1 pt] (noise) -- (n1);
	\end{tikzpicture}
	\caption{Two-user symbol-synchronous multiple access channel given in~\eqref{eq:channel output}.
	}
	\label{fig:MAC}
	\vspace{-0.2cm}
\end{figure}
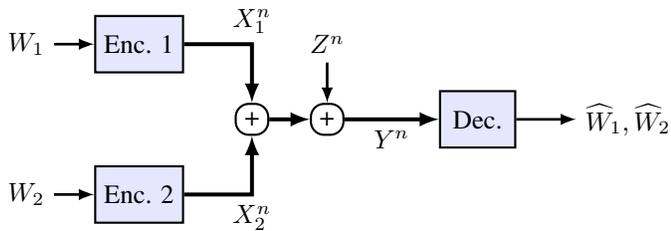
We study the  two-user, equal-power, symbol-synchronous GMAC with $L$-ary inputs, depicted in Fig. \ref{fig:MAC},
in which the channel output is obtained as
\ea{
	Y^n=X_1^n+X_2^n+Z^n,
	\label{eq:channel output}
}
where the additive noise $Z^n$ is an i.i.d.~sequence drawn from
$\Ncal(0,\sgs_z)$, the channel inputs at the two users, $X_i^n \in \Xcal_i^n$ $i\in \{1,2\}$, are subject to the power constraint\footnote{
	Note that the channel gains are absorbed into the power constraint in the model of \eqref{eq:channel output}, that is a channel with $Y'^n=h_1 X_1'^n+h_2 X_2'^n+Z^n,$ can be reduced to the model in \eqref{eq:channel output} without loss of generality.}
\ea{
\frac{1}{n}\sum_{i=1}^n X_{ki}^2 \leq  1, \quad k\in\{1,2\},
	\label{eq:sum constraint}
}
and the cardinality constraints
\ea{
	|\Xcal_k| \leq 2^L, \quad L \in \Nbb.
	\label{eq:card constraint}
}
For this model, the symmetric per-user capacity is obtained
\ea{
	& R_{\rm sym}= \max_{P_{X_1} P_{X_2}}
		\f12 I(Y ; X_1 , X_2),
	\label{eq:Rsym}
}
where the maximization is over all distributions $P_{X_k}$ with support $\Xcal_k$ and $\Ebb[X_k^2]=1$ for $k \in \{1,2\}$.
In the following, we consider the problem of jointly designing LDPC codes and constellations $\Xcal_1$ and $\Xcal_2$ to approach the symmetric capacity of the GMAC.
In particular, we consider the following architecture: (i) each transmitter produces $L$ LDPC codewords of length $n$, then (ii) each codeword is mapped to the input signal using $L$-level multilevel coding, and (iii) each bit-level is jointly decoded across the users while successive interference cancellation is applied across bit-levels.

\noindent
{\bf Multilevel Coding:}
We consider the scenario in which the sequences $X_k^n, \ k \in \{1,2\}$ are obtained by multilevel coding (MLC)~\cite{imai1977new} with $L$ bit-levels.
More precisely, we assume that \eqref{eq:card constraint} holds with equality so that
\ea{
	X_k^n= \sum_{i=1}^L U_{ki}^n, \quad k\in\{1,2\},
}
where $U_{ki}^n$ is a binary LDPC codeword  of length $n$ mapped onto the support $\Ucal_{ki}^n$ with $\Ucal_{ki}=\{l_{ki},h_{ki}\}$, where $l_{ki},h_{ki} \in \Rbb$ are the two levels of the modulation at bit-level $i$ for user $k$ (e.g., $l_{ki}={-1}$ and $h_{ki}={+1}$ for BPSK). The power of the bit-level $i$ is therefore $P_{ki}=l_{ki}^2+h_{ki}^2$ with $\sum_i^{L} P_{ki}=2L$. The rate of each bit-level is $R_{ki}$, so that the total attainable per-user rate is $\sum_i^L R_{ki}=R_k$.

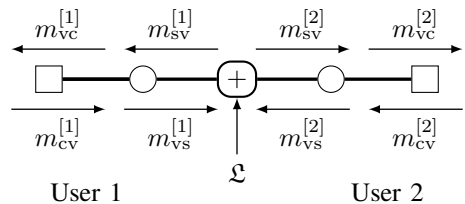
\begin{figure}[t]
	\centering
	\begin{tikzpicture}[node distance=1.25 cm,auto,>=latex]
	\node at (0,0) (MAC1) [rbox]{$+$};
	\node  [joint, right of=MAC1](v1m1){};
	\node  [joint, left of=MAC1](v2m1){};
	\draw[-,line width=1.250pt] (MAC1) -- (v1m1) ;
	\draw[-,line width=1.250pt] (MAC1) -- (v2m1) ;
	\node [check, right of=v1m1](c1m1){};
	\node [check, left  of=v2m1](c2m1){};
	\draw[-,line width=1.250pt] (c2m1) -- (v2m1) ;
	\draw[-,line width=1.250pt] (c1m1) -- (v1m1) ;
	\draw[-,line width=1.250pt] (c2m1) -- (v2m1) ;
	\draw[<-,line width=.5 pt] (-0.25,-.4) --node[below]{$m_{\rm vs}^{[1]}$}  (-1.5,-.4) ;
	\draw[->,line width=.5 pt] (-0.25,+.4) --node[above]{$m_{\rm sv}^{[1]}$} (-1.5,+.4) ;
	\draw[<-,line width=.5 pt] (0.25,-.4) --node[below]{$m_{\rm vs}^{[2]}$}  (+1.5,-.4) ;
	\draw[->,line width=.5 pt] (0.25,+.4) --node[above]{$m_{\rm sv}^{[2]}$} (+1.5,+.4) ;
	\draw[<-,line width=.5 pt] (-1.75,-.4) --node[below]{$m_{\rm cv}^{[1]}$}  (-3,-.4) ;
	\draw[->,line width=.5 pt] (-1.75,+.4) --node[above]{$m_{\rm vc}^{[1]}$} (-3,+.4) ;
	\draw[<-,line width=.5 pt] (+1.75,-.4) --node[below]{$m_{\rm cv}^{[2]}$}  (+3,-.4) ;
	\draw[->,line width=.5 pt] (+1.75,+.4) --node[above]{$m_{\rm vc}^{[2]}$} (+3,+.4) ;
	\node at (-2,-1.5) {User~1};
	\node at (+2,-1.5) {User~2};
	\draw [<-,line width=.5pt]  (MAC1)-- (0,-1);
	\node at (0,-1.25) {$\Lmf$};
	\end{tikzpicture}
	\caption{Messages employed in the MAC BP algorithm.}
	\label{fig:messages}
	\vspace{-0.2cm}
\end{figure}

\noindent
{\bf Multiuser Shaping:}
We refer to $\Xcal_k=\sum \Ucal_{ki}$ as the input constellation of user $k$, while the support of $X_{1,n}+X_{2,n}$ is denoted as $\Xcal_{\sum}$ and referred to as the \emph{sum constellation}. The input constellation of the users is chosen to maximize the the mutual information in~\eqref{eq:Rsym} under the power and cardinality constraints in~\eqref{eq:sum constraint} and \eqref{eq:card constraint}.
 We consider the case in which each coded bit is mapped to a MLC level, so that the input is uniformly distributed over the constellation points. Not that despite this, the constellation observed by the receiver is not necessarily uniformly distributed.

\begin{figure*}[t]
	\centering
	\small
	\begin{align}
	f^{[1,1]}(y,vs^{[2,1]}) & = \log \left(\frac{\sum _{b^{[1,2]},b^{[2,2]}}\mathbb{P}(y|b^{[1,1]}=0,b^{[2,1]}=0,b^{[1,2]},b^{[2,2]})e^{vs^{[2,1]}} + \sum _{b^{[1,2]},b^{[2,2]}}\mathbb{P}(y|b^{[1,1]}=0,b^{[2,1]}=1,b^{[1,2]},b^{[2,2]})}{\sum _{b^{[1,2]},b^{[2,2]}}\mathbb{P}(y|b^{[1,1]}=1,b^{[2,1]}=0,b^{[1,2]},b^{[2,2]})e^{vs^{[2,1]}} + \sum _{b^{[1,2]},b^{[2,2]}}\mathbb{P}(y|b^{[1,1]}=1,b^{[2,1]}=1,b^{[1,2]},b^{[2,2]})}\right) \label{eq:sv11}
	\end{align}
	\vspace{-0.1cm}
	\hrule
	\vspace{-0.1cm}
\end{figure*}

\noindent
{\bf Joint Per-Layer LDPC Decoding:}
For each bit-level $i$ of the MLC, the decoder jointly decodes the codewords transmitted by the two users, $\Ucal^n_{1i}$ and $\Ucal^n_{2i}$ using BP decoding and given the knowledge of the decoded bits from all previous bit-levels.
For each bit-level, BP decoding is performed on the factor graph in Fig.~\ref{fig:messages} where $\Lmf$ indicates the log-likelihood ratio of the channel output. The update rules for the variable-to-check messages, the check-to-variable messages, and the variable-to-state messages follow standard BP decoding rules.
The factor graph in Fig.~\ref{fig:messages} also contains a \emph{MAC state node}~\cite{roumy2007characterization}, which takes as inputs the channel output and the bit-reliability of one user and produces the bit-reliability of the other user.
As an example, the update rule for the state-to-variable message towards user $1$ in bit-level $1$ is given by $sv^{[1,1]} = f^{[1,1]}(y,vs^{[2,1]})$, where $f^{[1,1]}(y,vs^{[2,1]})$ is defined in~\eqref{eq:sv11}. An analogous expression can be obtained for update rule for the state-to-variable message towards user $2$ in bit-level $1$, i.e., $sv^{[2,1]} = f^{[2,1]}(y,vs^{[1,1]})$, as well as for all other bit-levels. We note that the channel LLR for user $1$ (resp. user $2$) in bit-level $i$ can be calculated as $f^{[1,i]}(y,0)$ (resp. $f^{[2,i]}(y,0)$).

\section{Input Constellation Optimization}
The constellation optimization is performed as follows. First the mutual information expression in \eqref{eq:Rsym} is maximized using a numerical optimization algorithm under the constraint that the input distribution is uniformly distributed over a set satisfying~\eqref{eq:sum constraint}. Then, the rate allocation for each bit-level is obtained by considering cancellation across layers. In particular, after the optimal constellation is determined, the rate of each bit-level is $R_i = I(Y;U_i|U^{i-1})$, where $i$ is the bit-level decoding order index.\footnote{Note that the overall rate remains the same whatever the order of decoding as long as correct decoding is guaranteed. This implies that the same codes can be reused for multiple SNR points by appropriately choosing their bit-level and the order of decoding.}

\section{LDPC Code Design}\label{sec:design}

Let $\lambda^{[k]}(x) = \sum _{j}\lambda^{[k]} _j x^{j-1}$ and $\rho^{[k]}(x) = \sum _{j}\rho^{[k]} _j x^{j-1}$ be the edge-perspective variable node and check node degree distribution polynomials for the LDPC code employed by user $k$, respectively. The node-perspective variable node degree distribution $L^{[k]}(x)$ is given by
\begin{align}
	L^{[k]}(x) & = \sum _{j} L^{[k]} _j x^{j} = \frac{\int _0^x \lambda^{[k]} (z)dz}{\int _0^1 \lambda^{[k]} (z)dz}.
\end{align}
The design rate of this code is
\begin{align}
	R^{[k]} & =  1 - \f{\sum_{j}\rho^{[k]}_j/j}{\sum_{j}\lambda^{[k]}_j/j}, \label{requ1}
\end{align}
where the bit-level $i$ is omitted for simplicity. As is common practice in the literature (e.g., \cite{chung2001}), we constrain $\rho^{[k]}(x)$ to be \emph{concentrated}, meaning that $\rho^{[k]}(x) = x^{d^{[k]}_c-1}, \, d^{[k]}_c \in \mathbb{N}$.

\subsection{EXIT Charts}
Density evolution (DE) can be used to analyze the decoding procedure over the GMAC. Extrinsic information transfer (EXIT) charts are a simpler analysis tool than DE that reduces the infinite-dimensional problem of tracking densities to a single-dimensional problem of tracking the mutual information between the decoder messages and the codeword bits. Moreover, EXIT charts enable the formulation of the LDPC code design as a linear program, which can be solved efficiently.

In the following, we temporarily omit the bit-level $i$ for simplicity of notation, since all expressions except~\eqref{eq:Isv} are common for all bit-levels. Let $I^{[k]}_{CV}$ (resp. $I^{[k]}_{SV}$) denote the average mutual information between the codeword bits and the check-to-variable (resp. state-to-variable) messages of user $k$. The EXIT chart $I^{j,[k]}_{VC}$ for the variable-to-check messages for variable node of degree $j$ is
\begin{align}
  I^{j,[k]}_{VC} & = J\left(\sqrt{(j-1)\left[J^{-1}(I^{[k]}_{CV})\right]^2 + \left[J^{-1}(I^{[k]}_{SV})\right]^2}\right),
\end{align}
where $J(\cdot)$ and $J^{-1}(\cdot)$ are given in \cite{tenbrink2004}. Averaging over $\lambda (x)$, we obtain the variable-to-check EXIT chart
\begin{align}
  I^{[k]}_{VC} & = \sum _i \lambda^{[k]}_j I^{j,[k]}_{VC}. \label{eqn:evc1}
\end{align}
Similarly, it can be shown that the EXIT chart $I^{[k]}_{VS}$ describing the variable-to-state messages is~\cite{roumy2007characterization,Balatsoukas2018}
\begin{align}
  I^{[k]}_{VS} & = \sum _j L^{[k]}_j J\left(\sqrt{j}J^{-1}(I^{[k]}_{CV})\right).
\end{align}
The EXIT chart describing the check-to-variable messages for user $k$ can be approximated as
\begin{align}
  I^{[k]}_{CV} & \approx 1 - J\left(\sqrt{(d^{[k]}_c-1)}J^{-1}(1-I^{[k]}_{VC})\right).
\end{align}
For the state-to-variable messages, we can make the all-zero codeword assumption for the user in question, but for the other user a typical codeword of type one-half has to be assumed~\cite{Balatsoukas2018}. As such, the average mutual information between the state-to-variable messages to user $1$ and this user's codeword bits is~\cite{roumy2007characterization,Balatsoukas2018}
\begin{align}
  I^{[1]}_{SV} & = \frac{1}{2}J\left( \sqrt{2F^{[1]}_{00}\left(m\right)} \right) + \frac{1}{2}J\left( \sqrt{2F^{[1]}_{01}\left(m\right)} \right), \label{eq:Isv}
\end{align}
where $F^{[1]}_{00}$ and $F^{[1]}_{01}$ denote the mean of the state-to-variable messages towards user $1$ given the distribution of the channel observation $y$ (see \cite{Balatsoukas2018}) and the mean $m$ of the (symmetric Gaussian distributed) variable-to-state messages from user $2$. The mean $m$ is given by
\begin{align}
	m & = \frac{1}{2}\left[J^{-1}\left(I^{[2]}_{VS}\right)\right]^2.
\end{align}
An analogous expression can be derived for user $2$. Since $F^{[1]}_{00}$ and $F^{[1]}_{01}$ are different for different bit-levels, at this point we re-introduce the bit-level index $i$. In general, we have
\begin{align}
	F^{[1,i]}_{0b}\left(m\right) & {=} \int_{\mathbb{R}^2} \mathbb{P}(y,vs^{[2,i]}|b^{[2,i]}{=}b) f^{[1,i]}(y,vs^{[2,i]}) \,dy\,dvs^{[2,i]}, \label{eq:F}
\end{align}
for $i\in\left\{1,\hdots,L\right\}$, and where $\mathbb{P}(y,vs^{[2,i]}|b^{[2,i]}) = \mathbb{P}(y|b^{[2,i]})\mathbb{P}(vs^{[2,1]}|b^{[2,i]})$. The distribution of the channel output $\mathbb{P}(y|b^{[2,i]})$ is generally a Gaussian mixture with means and weights of each component corresponding to the values and relative frequencies of the elements of $\Xcal_{\sum}$ for which $b^{[2,i]}{=}b$. Following a standard Gaussian approximation, we have $\mathbb{P}(vs^{[2,1]}|b^{[2,i]{=}0}) = \mathcal{N}(m,2m)$ and $\mathbb{P}(vs^{[2,1]}|b^{[2,i]{=}1}) = \mathcal{N}(-m,2m)$. Although simplified expressions for~\eqref{eq:F} have been derived in some special cases~\cite{roumy2007characterization,Balatsoukas2018}, in the general case we evaluate it using numerical integration.

\subsection{Degree Distribution Optimization}
Following standard arguments,  we argue that BP decoding is successful with high probability if the inverse of $I^{[k]}_{CV}$ lies below $I^{[k]}_{VC}$, for $k = 1,2$.
For our code design procedure, we set the maximum variable node degree to some $v_{\max}$. Typically, higher rates can be achieved with higher $v_{\max}$, but the decoding complexity of the LDPC code also increases. As in~\cite{Balatsoukas2018}, we fix the variable node degree distribution of one user and optimize the variable node degree distribution of the other user by alternately solving the following LP, for $k=1,2$
\eas{
  \text{maximize} \hspace{0.2cm} &  \sum _{i} \lambda ^{[k]}_j/j\\
  \text{subject to} \hspace{0.2cm} & I^{{-1},[k]}_{VC} < \sum _{j} \lambda ^{[k]} _j I^{j,[k]}_{VC}, \label{eq:exitconstr}\\
  & \sum _{i} \lambda ^{[k]}_j = 1, \quad j \in \left\{2,\hdots,v_{\max}\right\}, \\
  & \lambda ^{[k]}_j \geq 0, \quad\quad\;\;\, j \in \left\{2,\hdots,v_{\max}\right\},
%  & \sum _{i} \lambda ^{[k]}_j = 1,    \lambda ^{[k]}_j \geq 0, \;\, j = 2,\hdots,v_{\max},
}
where~\eqref{eq:exitconstr} is converted into multiple inequality constraints by discretization with step-size $\delta$.

\begin{table*}[t]
  \centering
  \scriptsize
  \setlength{\tabcolsep}{2.75pt}
  \caption{LDPC Code Optimization Results. $C$ denotes the per-level sum-capacity and $R$ denotes the rate of the designed LDPC codes.}\label{tab:results}
  \begin{tabular}{c|c|l|c|c|cccccccccccccccc}
    Const. & SNR &               & $C$                         & $R$        & $d_c$ & $\lambda_2$ & $\lambda_3$ & $\lambda_{34}$	& $\lambda_{35}$	& $\lambda_{37}$	& $\lambda_{38}$ 	& $\lambda_{39}$	& $\lambda_{46}$ & $\lambda_{50}$\\
    \hline
		\multirow{4}{*}{\rotatebox[origin=c]{0}{MC}} & \multirow{4}{*}{\rotatebox[origin=c]{0}{$10$~dB}} &
      User 1, level 1   &  \multirow{2}{*}{$1.0368$}  & $0.4953$   & $6$   & $0.3609$    & $0.4311$    & $0.1771$				& $0.0309$ 				& 				 				&									& 								& 								&\\
    && User 2, level 1   &  				  									& $0.4954$   & $6$   & $0.3609$    & $0.4312$    & $0.1749$				& $0.0330$ 				& 								& 								& 								& 								&\\
    && User 1, level 2   &  \multirow{2}{*}{$1.1106$}  & $0.5237$   & $6$   & $0.4377$    & $0.3786$    & 								&  								& $0.1077$				& $0.0760$ 				& 								& 								&\\
    && User 2, level 2   &  	  												& $0.5238$   & $6$   & $0.4377$    & $0.3787$    & 								&  								& $0.0960$				& $0.0876$ 				& 								& 								&\\
		\hline
		\multirow{4}{*}{\rotatebox[origin=c]{0}{MC}} & \multirow{4}{*}{\rotatebox[origin=c]{0}{$18$~dB}} &
      User 1, level 1   &  \multirow{2}{*}{$1.1554$}  & $0.5237$   & $6$   & $0.4132$    & $0.4174$    & 								&  								& 								& $0.0370$ 				& $0.1324$ 				& 								&\\
    && User 2, level 1   &    													& $0.5237$   & $6$   & $0.4132$    & $0.4174$    & 								&  								& 								& $0.0374$ 				& $0.1320$ 				& 								&\\
    && User 1, level 2   &  \multirow{2}{*}{$1.4988$}  & $0.7109$   & $10$  & $0.6779$    &     				 & 								&  								& 								&  								&  								& $0.3221$ \\
    && User 2, level 2   &   													& $0.7109$   & $10$  & $0.6779$    &     				 & 								&  								& 								&  								&   							& $0.3221$ \\
    \hline
		\multirow{4}{*}{\rotatebox[origin=c]{0}{Opt}} & \multirow{4}{*}{\rotatebox[origin=c]{0}{$18$~dB}} &
      User 1, level 1   &  \multirow{2}{*}{$1.3294$}  & $0.6414$   & $8$   & $0.4633$    & $0.3390$    & 								&  								& 								&  								&  								& 						& $0.1977$ \\
    && User 2, level 1   &    													& $0.5650$   & $8$   & $0.3498$    & $0.3175$    & 								&  								& 								&  								&  								& 						& $0.3328$ \\
    && User 1, level 2   &  \multirow{2}{*}{$1.9880$}  & $0.9000$   & $20$  & $1.0000$    &     				 & 								&  								& 								&  								&  								&  						& \\
    && User 2, level 2   &   													& $0.8732$   & $20$  & $0.3660$   & $0.6340$		 & 								&  								& 								&  								&   							&  						& \\
    \hline
  \end{tabular}
	\vspace{-0.2cm}
\end{table*}

\section{Simulation Results}
In this section, we present results for the proposed constellation and LDPC code design methods with $L=2$ bit-levels. All simulated LDPC codes have a blocklength of $n = 10,000$ and are constructed using the PEG algorithm~\cite{Hu2005}.

\begin{figure}[t]
	\centering
	\begin{tikzpicture}

	\pgfplotsset{grid style={dashed}}
	\small
	
	\begin{axis}[
		width = 0.95\columnwidth,
		height = 0.7\columnwidth,
		xlabel = {SNR (dB)},
		ylabel = {Sum-capacity (bits/ch. use)},
		ylabel near ticks,
		xlabel near ticks,
		xtick distance=4,
		%ytick distance=1,
		%label style={font=\footnotesize},
		%tick label style={font=\footnotesize},
		xmin = 6, xmax = 30,
		ymin = 0, ymax = 5,
		grid = both,
		%ymajorgrids,
		%legend pos = south east,
		%legend style={font=\scriptsize},
		%legend columns=1,
		%legend cell align=left,
		%legend entries={Gaussian, Max. Collisions, Superposition, Optimized}
	]		

		\addplot[black, thick, solid] table[x index = 0, y index = 1, y expr=2*\thisrowno{1}] {fig/data/Rmac.dat}; \label{p1}
		\addplot[blue, thick, solid, mark=*] table[x index = 0, y index = 1, y expr=2*\thisrowno{1}] {fig/data/Rmaxcol.dat}; \label{p2}
		\addplot[red, thick, solid, mark=square*] table[x index = 0, y index = 1, y expr=2*\thisrowno{1}] {fig/data/Rsup.dat}; \label{p3}
		\addplot[green!75!black, dashed, thick, mark=triangle*, mark options={scale=1.5, solid}] table[x index = 0, y index = 1, y expr=2*\thisrowno{1}] {fig/data/Ropt.dat}; \label{p4}

		\addplot[black, thick, mark=*, mark options={scale=1.25,fill=yellow}] coordinates {(10, 2.0380)}; \label{p5}
		\addplot[black, thick, mark=*, mark options={scale=1.25,fill=yellow}] coordinates {(18, 2.4692)}; \label{p6}
		\addplot[black, thick, mark=triangle*, mark options={scale=1.5,fill=yellow}] coordinates {(18, 2.9796)}; \label{p7}

	\end{axis}

	\node [draw,fill=white] at (rel axis cs: 0.53,0.19) {\scriptsize \shortstack[l]{
	\ref{p1} Gaussian \\
	\ref{p2} Max. Collisions \\
	\ref{p3} Superposition \\
	\ref{p4} Optimized}};

	% Second "Legend" node
	\node [draw,fill=white] at (rel axis cs: 0.03,0.875) {\scriptsize \shortstack[l]{
	\ref{p5} MC LDPC Sum-Rate \\
	\ref{p7} Opt LDPC Sum-Rate}};

\end{tikzpicture}%
	\caption{Sum-capacity of the two-user GMAC with Gaussian, MC, SP, and optimized inputs. Yellow points show the sum-rates of designed LDPC codes.}
	\label{fig:rates}
	\vspace{-0.2cm}
\end{figure}
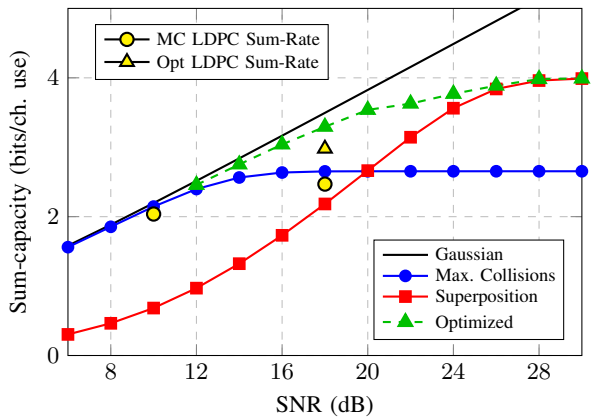

\subsection{Constellation Design}
A fundamental intuition about the 2-user GMAC with $L$-ary input capacity behavior is that the low-SNR performance is dominated by the shaping gain of the received constellation while the high-SNR behavior is dominated by the minimum distance between the points in the sum constellation.
For this reason, we consider two constellations in the following: the maximum-collisions (MC) and the superposition (SP) constellation, defined as
\eas{
	\mathcal{X}^{\text{MC}}_1	& = \left\{{-}1.342, {-}0.447, {+}0.447, {+}1.342 \right\}, \\
	\mathcal{X}^{\text{MC}}_2	& = \left\{{-}1.342, {-}0.447, {+}0.447, {+}1.342 \right\},
}
and
\eas{
	\mathcal{X}^{\text{SP}}_1	& = \left\{{-}0.335, {-}0.112, {+}0.112, {+}0.335 \right\}, \\
	\mathcal{X}^{\text{SP}}_2	& = \left\{{-}1.342, {-}0.447, {+}0.447, {+}1.342 \right\},
}
respectively. The MC constellation results in a  discrete triangular distribution of the received constellation which provides a shaping gain at low SNR.
The SP constellation, instead, maximizes the minimum distance between points in the sum constellation.
By numerically optimizing the sum-constellation we observe that significant gains can be achieved in terms of the sum-capacity in the region where both the MC and SP constellation pairs are far from optimal, i.e., for SNRs between $12$~dB and $24$~dB, as shown in Fig.~\ref{fig:rates}.
The numerically optimal constellation in this rate regime is approximatively equal to
\eas{
	\mathcal{X}^{\text{Opt}}_1	& = \left\{{-}1.316, {-}0.519, {+}0.519, {+}1.316 \right\}, \\
	\mathcal{X}^{\text{Opt}}_2	& = \left\{{-}1.406, {-}0.150, {+}0.150, {+}1.406 \right\}.
}{\label{eq:mid opt}}

\begin{figure}[t]
	\centering
	\begin{tikzpicture}

	\pgfplotsset{grid style={dashed}}
	\small

	\begin{semilogyaxis}[
		width = 0.95\columnwidth,
		height = 0.7\columnwidth,
		xlabel = {SNR (dB)},
		ylabel = {Bit Error Rate},
		ylabel near ticks,
		xlabel near ticks,
		xmin = 9.95, xmax = 22,
		ymin = 1e-6, ymax = 3e-1,
		grid = both,
		legend style={at={(0.485,-0.2)},anchor=north,font=\scriptsize},
		legend cell align={left},
		legend columns={2},
		transpose legend,
	]

		\addplot[blue, thick, solid, mark=*, mark options={scale=0.8}] table[x index=0, y index = 1] {fig/data/RES_n_10000_vmax_50_designSNR_10.00_pegH.dat};
		\addlegendentry{MC/MC (DSNR = $10$ dB)}
	
		\addplot[red, thick, solid, mark=square*, mark options={scale=0.8}] table[x index=0, y index = 1] {fig/data/RES_n_10000_vmax_50_designSNR_18.00_pegH.dat};
		\addlegendentry{MC/MC (DSNR = $18$ dB)}

		\addplot[magenta, thick, solid, mark=pentagon*, mark options={scale=0.9}] table[x index=0, y index = 1] {fig/data/RES_n_10000_vmax_50_designSNR_18.00_pegH_optConst.dat};
		\addlegendentry{MC/Opt (DSNR = $18$ dB)}

		\addplot[green!75!black, thick, solid, mark=triangle*, mark options={scale=1}] table[x index=0, y index = 1] {fig/data/RES_n_10000_vmax_50_designSNR_18.00_pegHoptConstDesign.dat};
		\addlegendentry{Opt/Opt  (DSNR = $18$ dB)}

		% Design design SNR lines
    \addplot +[black, solid, ultra thick, mark=none] coordinates {(10, 1e-7) (10, 1e-5)};
    \addplot +[black, solid, ultra thick, mark=none] coordinates {(18, 1e-7) (18, 1e-5)};
    %\addplot +[black, solid, ultra thick, mark=none] coordinates {(4.7712, 5e-7) (4.7712, 1e-4)};

		% Add annotations to design SNR lines
		%\node[anchor=west] (source1) at (axis cs:10,1.3e-5){\scriptsize $\text{SNR}{=}10.00$~dB};
		%\node (destination1) at (axis cs:10.04,1.1e-4){};
		%\node[anchor=west] (source2) at (axis cs:18,1.5e-4){\scriptsize $\text{SNR}{=}18$~dB};
		%\node (destination2) at (axis cs:18,9e-5){};
		%\node[anchor=west] (sourceP115) at (axis cs:0.7,4e-4){\scriptsize $\text{SNR}_1{=}1.76$~dB};
		%\node (destinationP115) at (axis cs:1.75,9e-5){};

		% Add arrows
		%\draw[-{Latex[length=1.75mm,width=1.25mm]}](source1)--(destination1);
		%\draw[-{Latex[length=1.75mm,width=1.25mm]}](source2.east)--(destination2);
		%\draw[-{Latex[length=1.75mm,width=1.25mm]}](sourceP115.south)--(destinationP115);

	\end{semilogyaxis}%
\end{tikzpicture}%
\vspace{-0.025in}
	\caption{Average (over the two users and the two bit-layers) finite-length BER performance of LDPC codes designed for different constellations. The thick black lines correspond to the two considered design SNRs.}
	\label{fig:bervssnr}
	\vspace{-0.2cm}
\end{figure}
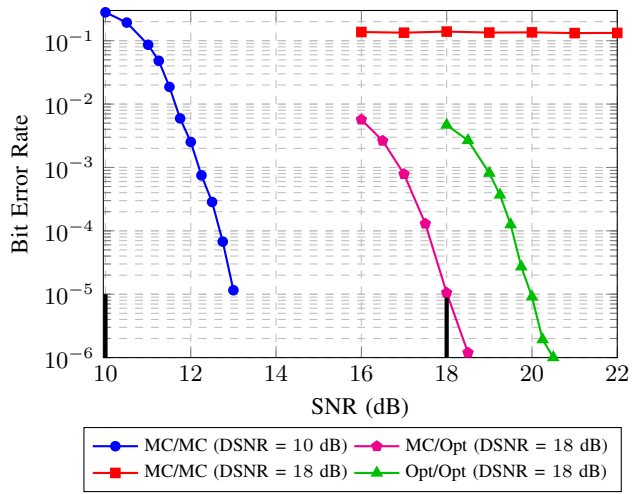

\subsection{LDPC Code Design}
In order to verify the LDPC code design procedure described in Section~\ref{sec:design}, we first design and evaluate LDPC codes for an MC constellation at a design SNR (DSNR) of $10$~dB. The resulting degree distributions for each user and each bit-level are given in Table~\ref{tab:results}. The maximal sum-rate of the two-user GMAC with an MC constellation at an SNR of $10$~dB is $2.1474$ bits per channel use (\bpcu) and is illustrated in Fig.~\ref{fig:rates}. The sum-rate of the designed LDPC codes is $2.0380$ \bpcu, meaning that we are only $0.1094$ \bpcu away from the maximal sum-rate. The finite-length performance of the designed LDPC codes is shown in Fig.~\ref{fig:bervssnr}, where the bit-error rate has been averaged over the two users and over the two bit-levels. The ``$X$/$Y$'' notation in the legend means that the LDPC codes are designed for constellation $X$ and constellation $Y$ is used.
In Table~\ref{tab:results}, we also give degree distributions for a DSN of $18$~dB, where the maximal sum-rate of the two-user GMAC with an MC constellation is $2.6542$ \bpcu. The sum-rate of the designed LDPC codes is $2.4692$ \bpcu, meaning that we are $0.185$ \bpcu away from the maximal sum-rate. Unfortunately, the MC/MC finite-length simulation results in Fig.~\ref{fig:bervssnr} reveal that, in this case, the gap to capacity is not sufficient to make up for the performance loss due to finite-length effects of the LDPC codes. In fact, because we are operating in the inter-user interference limited regime, increasing the SNR does not improve the BER performance.

\subsection{Joint Constellation \& LDPC Code Design}
In Fig.~\ref{fig:bervssnr}, we observe that by simply replacing the MC constellation with the constellation in \eqref{eq:mid opt} and using the LDPC codes designed for the MC constellation (MC/Opt) at a DSNR of $18$~dB dramatically improves the performance compared to the MC/MC case. This is not unexpected, since the maximal sum-rate of the two-user GMAC with the optimized constellation is significantly higher than when using the MC constellation (cf. Fig.~\ref{fig:rates}), so that the gap of the LDPC code sum-rate to capacity is relatively large.

We also designed LDPC codes specifically for the optimized constellation using the code design procedure described in Section~\ref{sec:design}. The corresponding degree distributions are given in Table~\ref{tab:results} for a DSNR of $18$~dB. The maximal sum-rate of the two-user GMAC with the optimized constellation at an SNR of $18$~dB is $3.3174$ \bpcu, while the sum-rate of the LDPC codes is $2.9796$~\bpcu. In Fig.~\ref{fig:bervssnr}, we observe that, contrary to the MC/MC case where decoding fails at all SNRs, the optimized constellation with appropriately designed LDPC codes achieves a BER of $10^{-6}$ at a $2.5$~dB from the design SNR, while also having a significantly higher sum-rate.

\subsection{Discussion}
For the optimized constellation, the rate for bit-level $2$ is close to $1$ \bpcu, making the design of good binary codes challenging. It may be useful to modify the constellation design in order to balance the per-level rates, possibly sacrificing some of the multi-user shaping gain in the process. Another important issue is the practical scalability of the scheme to more users and higher order constellations, as in its current form the scheme requires different codes for each user and for each bit-level. Moreover, the state node becomes increasingly complex as more users are added to the system. Finally, we currently assume that the channel gains are fixed, but the code design should be extended to fading channels.

\section{Conclusion}
In this work, we jointly designed optimal higher-order constellations and LDPC codes for the two-user GMAC. We showed that, by optimizing the constellations that the users employ, we can obtain significant multi-user shaping gains of up to $0.88$ \bpcu, which can be harnessed efficiently using per-user jointly decoded MLC with LDPC codes.

\section{Acknowledgment}
The work of Alexios Balatsoukas-Stimming is supported by the SNSF project \#175813. The work of S. Rini is supported by the MOST grant 107-2221-E-009-032-MY3. The work of Joerg Kliewer is supported by NSF Grant CCF-1815322. The authors would like to thank the anonymous reviewers for their useful suggestions.

\bibliographystyle{IEEEtran}
\bibliography{IEEEabrv,steBib1}

\end{document}